\documentclass[prl,twocolumn,superscriptaddress,10pt,floatfix]{revtex4}
\usepackage[pdftex,plainpages=false,colorlinks=true,citecolor=blue,linkcolor=blue,urlcolor=blue,filecolor=green,bookmarksopen=true]{hyperref}
\usepackage[utf8]{inputenc}
\usepackage{amsmath}
\usepackage[caption = false]{subfig}
\usepackage{graphicx,epstopdf}
\usepackage{blindtext}
\usepackage{lipsum}
\usepackage{amsfonts}
\usepackage{bbm}
\usepackage{amssymb}
\usepackage{enumerate}
\usepackage{color}
\usepackage{latexsym}
\usepackage{times,txfonts}
\newtheorem{theorem}{Theorem}

\newcommand{\Ncal}{\mathcal{N}}

\newcommand{\Fcal}{\mathcal{F}}
\newcommand{\Ocal}{\mathcal{O}}

\DeclareFontFamily{U}{mathc}{}
\DeclareFontShape{U}{mathc}{m}{it}%
{<->s*[1.03] mathc10}{}
\DeclareMathAlphabet{\mathscr}{U}{mathc}{m}{it}
\newcommand{\Ocalb}{\mathscr{O}}

\newcommand{\ket}[1]{| #1 \rangle}

\newcommand{\interpro}[2]{\langle #1 | #2 \rangle}
\newcommand{\bra}[1]{\langle #1 |}

\usepackage{soul}

\begin{document}

\title{Validation of Quantum Adiabaticity through Non-Inertial Frames and Its Trapped-Ion Realization}

\author{Chang-Kang Hu}
\affiliation{CAS Key Laboratory of Quantum Information, University of Science and Technology of China, Hefei, 230026, People’s Republic of China}
\affiliation{Synergetic Innovation Center of Quantum Information and Quantum Physics,University of Science and Technology of China, Hefei, 230026, People’s Republic of China}

\author{Jin-Ming Cui}
\email{jmcui@ustc.edu.cn}
\affiliation{CAS Key Laboratory of Quantum Information, University of Science and Technology of China, Hefei, 230026, People’s Republic of China}
\affiliation{Synergetic Innovation Center of Quantum Information and Quantum Physics,University of Science and Technology of China, Hefei, 230026, People’s Republic of China}

\author{Alan C. Santos}
\email{ac\_santos@id.uff.br}
\affiliation{Instituto de F\'{i}sica, Universidade Federal Fluminense, Av. Gal. Milton Tavares de Souza s/n, Gragoat\'{a}, 24210-346 Niter\'{o}i, Rio de Janeiro, Brazil}

\author{\\Yun-Feng Huang}
\email{hyf@ustc.edu.cn}
\affiliation{CAS Key Laboratory of Quantum Information, University of Science and Technology of China, Hefei, 230026, People’s Republic of China}
\affiliation{Synergetic Innovation Center of Quantum Information and Quantum Physics,University of Science and Technology of China, Hefei, 230026, People’s Republic of China}

\author{Chuan-Feng Li}
\email{cfli@ustc.edu.cn}
\affiliation{CAS Key Laboratory of Quantum Information, University of Science and Technology of China, Hefei, 230026, People’s Republic of China}
\affiliation{Synergetic Innovation Center of Quantum Information and Quantum Physics,University of Science and Technology of China, Hefei, 230026, People’s Republic of China}

\author{Guang-Can Guo}
\affiliation{CAS Key Laboratory of Quantum Information, University of Science and Technology of China, Hefei, 230026, People’s Republic of China}
\affiliation{Synergetic Innovation Center of Quantum Information and Quantum Physics,University of Science and Technology of China, Hefei, 230026, People’s Republic of China}

\author{Frederico Brito}
\email{fbb@ifsc.usp.br}
\affiliation{Instituto de F\'{i}sica de S\~ao Carlos, Universidade de S\~ao Paulo, C.P. 369, S\~ao Carlos, SP, 13560-970, Brazil}

\author{Marcelo S. Sarandy}
\email{msarandy@id.uff.br}
\affiliation{Instituto de F\'{i}sica, Universidade Federal Fluminense, Av. Gal. Milton Tavares de Souza s/n, Gragoat\'{a}, 24210-346 Niter\'{o}i, Rio de Janeiro, Brazil}

\begin{abstract}
Validity conditions for the adiabatic approximation are useful tools to understand and predict the quantum dynamics.
Remarkably, the resonance phenomenon in oscillating quantum systems has challenged the adiabatic theorem.
In this scenario, inconsistencies in the application of quantitative adiabatic conditions have led to a sequence of
new approaches for adiabaticity. Here, by adopting a different strategy, we introduce a validation mechanism
for the adiabatic approximation by driving the quantum system to a non-inertial reference frame. More specifically, we
begin by considering several relevant adiabatic approximation conditions previously derived and show that all of
them fail by introducing a suitable oscillating Hamiltonian for a single quantum bit (qubit). Then, by evaluating the adiabatic
condition in a rotated non-inertial frame, we show that all of these conditions, including the standard adiabatic condition,
can correctly describe the adiabatic dynamics in the original frame, either far from resonance or at a resonant point.
Moreover, we prove that this validation mechanism can be extended for general multi-particle quantum systems, establishing
the conditions for the equivalence of the adiabatic behavior as described in inertial or non-inertial frames.
In order to experimentally investigate our method, we consider a hyperfine qubit through a single trapped
Ytterbium ion $^{171} $Yb$^+$, where the ion hyperfine energy levels are used as degrees of freedom of a two-level system.
By monitoring the quantum evolution, we explicitly show the consistency of the adiabatic conditions
in the non-inertial frame.

\end{abstract}

\maketitle

The adiabatic theorem~\cite{Born:28,Kato:50,Messiah:Book} is a fundamental ingredient in a number of applications in quantum mechanics.
Under adiabatic dynamics, a quantum system evolves obeying a sufficiently slowly-varying Hamiltonian, which prevents
changes in the populations of the energy eigenlevels. In particular, if the system is prepared in an eigenstate $|E_n(0)\rangle$ of the
Hamiltonian $H(t)$ at a time $t=0$, it will evolve to the corresponding instantaneous eigenstate $|E_n(t)\rangle$ at later times.
The concept of adiabaticity plays a relevant role in a vast array of fields, such as energy-level crossings in molecules \cite{Landau:32,Zener:32},
quantum field theory~\cite{Gellmann:51}, geometric phases~\cite{Berry:84,Wilczek:84}, quantum computation~\cite{Farhi:01,Bacon:09,Hen:15,Santos:16,Tameem-Lidar:Book},
quantum thermodynamics~\cite{Abah:17,He:02}, among others. However, despite such a wide range of applications,
both \textit{sufficiency} and \textit{necessity} of quantitative conditions for the adiabatic behavior have been challenged~\cite{Marzlin:04}.
In particular, inconsistencies in the application of the adiabatic theorem may appear for oscillating Hamiltonians as a consequence of
resonant transitions between their energy levels~\cite{Duki:05,Amin:09}. Such inconsistencies have led to a revisitation of the adiabatic theorem,
yielding many new proposals of adiabatic conditions (ACs) in more general settings (see, e.g., Refs.~\cite{Ambainis:04,Tong:07,Jian-da-wu:07,Jianda-Wu:08}).

The first experimental investigation on the comparison among these proposals has been considered by Du \textit{et al.}~\cite{Suter:08},
where the authors considered a single nuclear spin-1/2 particle in a rotating magnetic field manipulated by nuclear magnetic resonance (NMR) techinques.
It is then shown the violation of both the sufficiency and necessity of the traditional AC, with partial success via some other generalized ACs.
It is remarkable such possible violations are already apparent for a single quantum bit (qubit) system. More specifically,
the ACs analyzed in Ref.~\cite{Suter:08} can be cast in the form of \textit{adiabatic coefficients} $C_{n}(t)$ which, for a qubit, are given by
\begin{subequations}
\begin{eqnarray}
C_{1} &=& \max_{t \in [ 0,\tau ]} \left|\frac{|\bra{E_{0}(t)} \dot{H}(t)\ket{E_{1}(t)}|}{\left[ E_{0}(t) - E_{1}(t) \right]^2}\right| , \label{Cond1}\\
C_{2} &=& \max_{t \in [ 0,\tau ]} \left| \frac{d}{dt}\left(\frac{\bra{E_{0}(t)} \dot{H}(t)\ket{E_{1}(t)}}{\left[ E_{0}(t) - E_{1}(t) \right]^2}\right) \right| \tau , \label{Cond2} \\
C_{3} &=& \max_{t \in [ 0,\tau ]}  \left| \frac{\left \vert d_{10}(t) \right \vert}{| E_{1}(t) - E_{0}(t) - \Delta_{10}(t)|} \right| , \label{Cond3} \\
C_{4} &=& \max_{t \in [ 0,\tau ]}
\left\{ \frac{\tau^2 ||\dot{H}(t)||^3}{|E_{0}(t) - E_{1}(t)|^4} , \frac{\tau^2 ||\dot{H}(t)||\cdot||\ddot{H}(t)||}{|E_{0}(t) - E_{1}(t)|^3}
\right\}  , \label{Cond4}
\end{eqnarray}
\label{Conditions}
\end{subequations}
\hspace{-0.15cm}where $\ket{E_{n}(t)}$ are eigenstates of $H(t)$ with energies $E_{n}(t)$, $\tau$ is the total evolution time, $d_{10}(t) = \frac{\bra{E_{1}(t)}\dot{H}(t)\ket{E_{0}(t)}}{E_{0}(t)-E_{1}(t)}$, $
\Delta_{10}(t) = i\gamma_{1}(t) - i\gamma_{0}(t) + \frac{d }{dt} \arg [i d_{10}(t)]$, and $\gamma_{n}(t) = \interpro{E_{n}(t)}{\dot{E}_{n}(t)}$, with the {\it dot} symbol denoting
time derivative and $|| \cdot ||$ denoting the usual operator norm. The adiabaticity coefficient $C_{1}$ is the well-known \textit{standard (traditional)} adiabatic condition \cite{Messiah:Book,Amin:09,Sarandy:04}, while the conditions $C_{2}$, $C_{3}$ and $C_{4}$ as shown above were derived by Tong \textit{et al} \cite{Tong:07}, Wu \textit{et al} \cite{Jian-da-wu:07,Jianda-Wu:08} and Ambainis-Regev \cite{Ambainis:04}, respectively. In general, the adiabatic behavior in a quantum system is achieved when $C_{n} \ll 1$. In Ref.~\cite{Suter:08}, it is shown that the condition $C_{1}$ is neither sufficient nor necessary for guaranteeing the adiabatic behavior, while the conditions $C_{2}$, $C_{3}$, and $C_{4}$ seem to successfully indicate the resonant phenomena observed in their specific experiment (even though their validity is debatable for more general quantum systems).

Here, instead of looking for new proposals of ACs, we adopt a different strategy to analyze the conditions in Eq.~\eqref{Conditions}.
More specifically, we consider the dynamics of the system in a non-inertial reference frame, where it is possible to show that all the conditions in Eq.~\eqref{Conditions},
including the traditional AC, work with no violations with respect to the exact solution of Sch\"odinger equation. To this end, we consider a
Hamiltonian that leads to a failure of all coefficients in Eq.~\eqref{Conditions}, which means that all the ACs in Eq.~\eqref{Conditions} are
neither necessary nor sufficient to describe the adiabaticity of the system. Remarkably, by implementing a change of reference frame, all of ACs
become both necessary and sufficient conditions for the example considered. We then generalize our results to a generic many-body Hamiltonian,
where we provide conditions for the equivalence of the adiabatic behavior in both inertial and non-inertial reference frames
These theoretical results are
realized in a single trapped Ytterbium ion $^{171} $Yb$^+$ system, with excellent experimental agreement.

\emph{Oscillating Hamiltonian for a single qubit in trapped ions: ACs in a non-inertial reference frame --} Let us begin by considering the Hamiltonian
\begin{eqnarray}
H(t) = \omega_{0} \sigma_{z} + \omega_{\text{T}} \sin(\omega t)\sigma_{x}  \text{ , } \label{ResH}
\end{eqnarray}
where we assume $|\omega_{0}|\gg |\omega_{\text{T}}|$. The set of eigenvectors of $H(t)$ is given by $\ket{E_{n}(t)} = \Ncal^{-1}_{n}(t)\left[-(-1)^n\alpha_{n}(t)\ket{0} + \ket{1}\right]$, $\Ncal_{n}^{2}(t) = 1+ \alpha_{n}^2(t)$, and $
\alpha_{n}(t) = \frac{1}{2}\cos \theta \csc (\omega t)\left[ -2(-1)^{n} \cos\theta + \Sigma  \right]$, with $\Sigma^2 = 3+\cos(2\theta)-2\cos(2\omega t)\sin^2 (\theta)$, $\theta = \arctan(\omega_{0}/\omega_{\text{T}})$, and $n\in \{0,1\}$. The energies are $E_{n}(t) = -(-1)^n \omega_{0}\Sigma/2$. Here we describe the system dynamics by
\begin{equation}
\dot{\rho}(t) = (-i/\hbar) [H(t),\rho(t)] \text{ . } \label{VonNeuEq}
\end{equation}

\begin{figure}[t]
\centering
\includegraphics[scale=0.271]{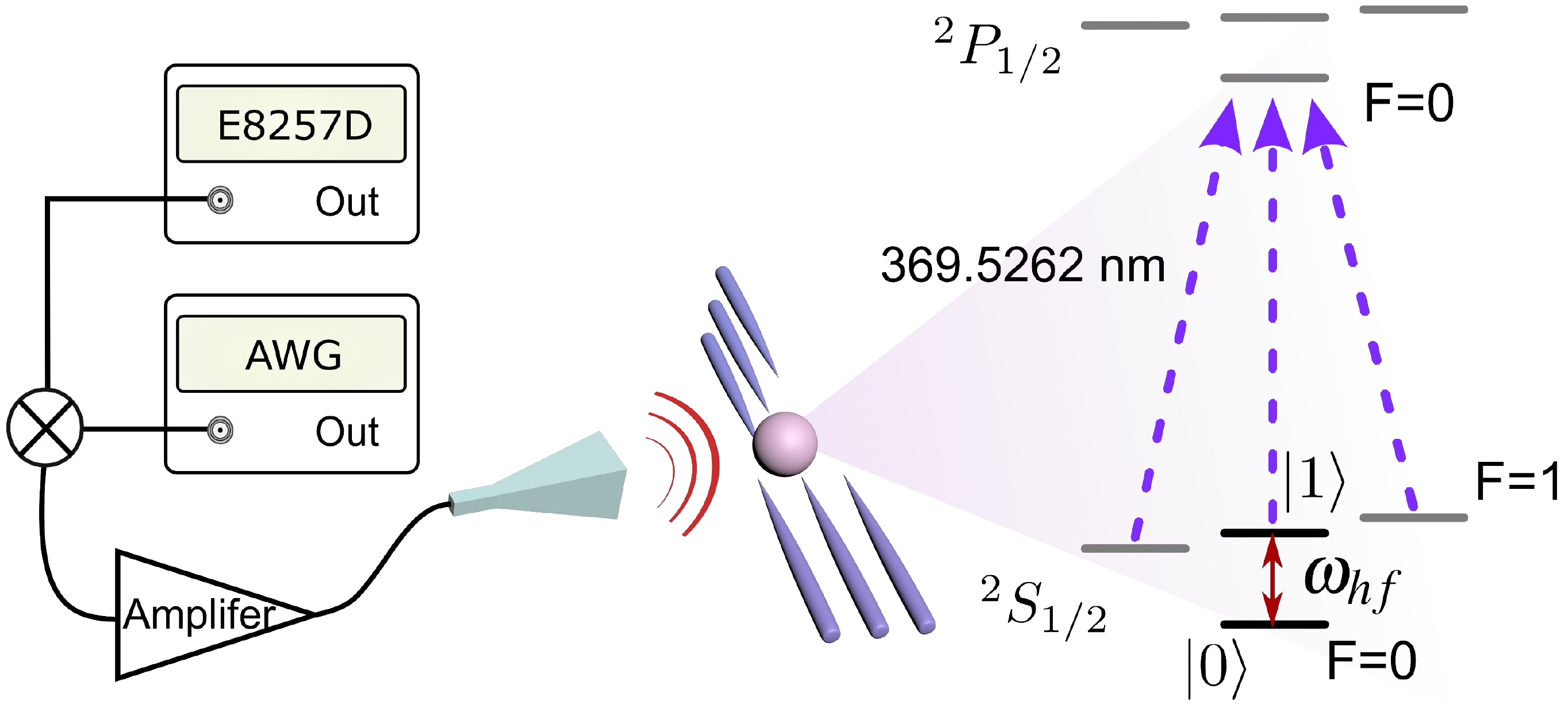}
\caption{Experimental setup for the validation of the adiabatic dynamics through non-inertial frames. The single $^{171}$Yb$^+$ ion is trapped in a six-needle Paul trap. 
The qubit is encoded in the hyperfine energy levels and coherently driven with a programmable AWG. The 369.5 nm laser is used to implement the quantum state detection.}\label{setup}
\end{figure}

 The experiment is performed using a single Ytterbium ion $^{171}$Yb$^+$, which is trapped in a six needle Paul trap, with the experimental setup schematically shown in Fig.~\ref{setup}.  The qubit is encoded in the hyperfine energy levels of $^{171}$Yb$^+$ (a hyperfine qubit), represented as ${\ket{0} \equiv \,^{2}S_{1/2}\, \ket{F=0,m_{F}=0}}$ and ${\ket{1} \equiv \,^{2}S_{1/2}\, \ket{F=1,m_{F}=0}}$~\cite{Olmschenk:07}. We coherently drive the hyperfine qubit with a programmable arbitrary waveform generator (AWG) \cite{Hu:18} after Doppler cooling and standard optical pumping process. A 369.5 nm laser is used for fluorescence detection to measure the population of the $\ket{1}$ state. Observation of more than one photon implies population in $\ket{1}$.

The system is initialized in the state $\ket {0}$ with optical pumping, so that the adiabatic dynamics is achieved if the system evolves as $\ket{\psi_{\text{ad}}(t)} = \ket{E_{1}(t)}$, up to a global phase. It is possible to show that the Hamiltonian presents a near-to-resonance situation when we set $|\omega - \omega_{0}| \ll |\omega_{\text{T}}|$. Thus, to study the adiabaticity validity conditions in our Hamiltonian in Eq.~\eqref{ResH} we need to compute the coefficients in Eq.~\eqref{Conditions} for different values of the $\omega$. In our experiment, we set the detuning $\omega_{0} = 2\pi \times 1.0$ MHz, the coupling strength $\omega_{\text{T}} = 2\pi \times 20.0$ KHz, and $\omega = a \times \omega_{0}$ ($a =10.0, 1.0173, 1.0, 0.9827~{\rm and}~ 0.1$, respectively).

\begin{figure*}
	\centering
	\subfloat[Fidelity]{\includegraphics[scale=0.245]{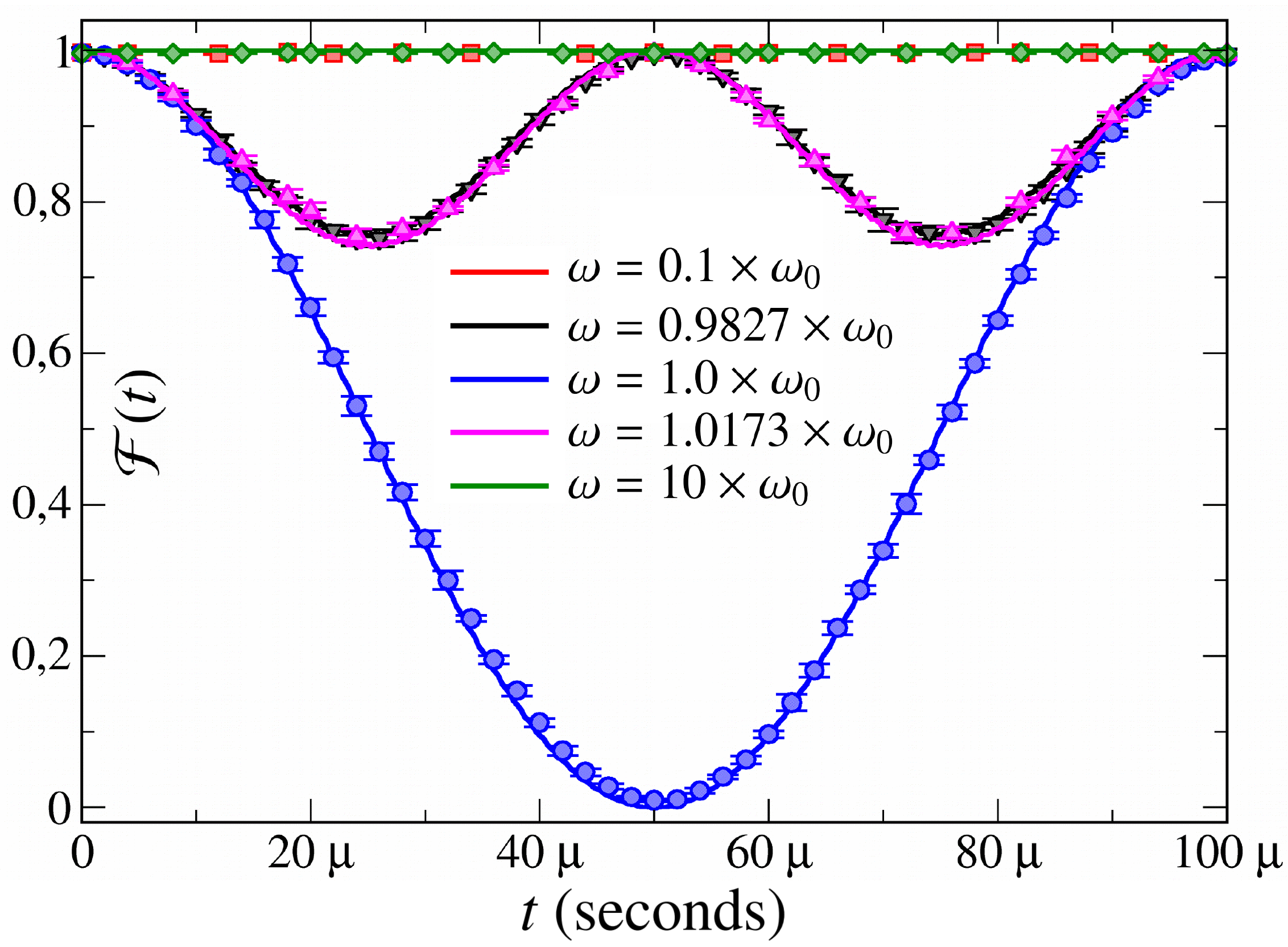}\label{Fig1}}\quad
	\subfloat[Inertial frame]{\includegraphics[scale=0.25]{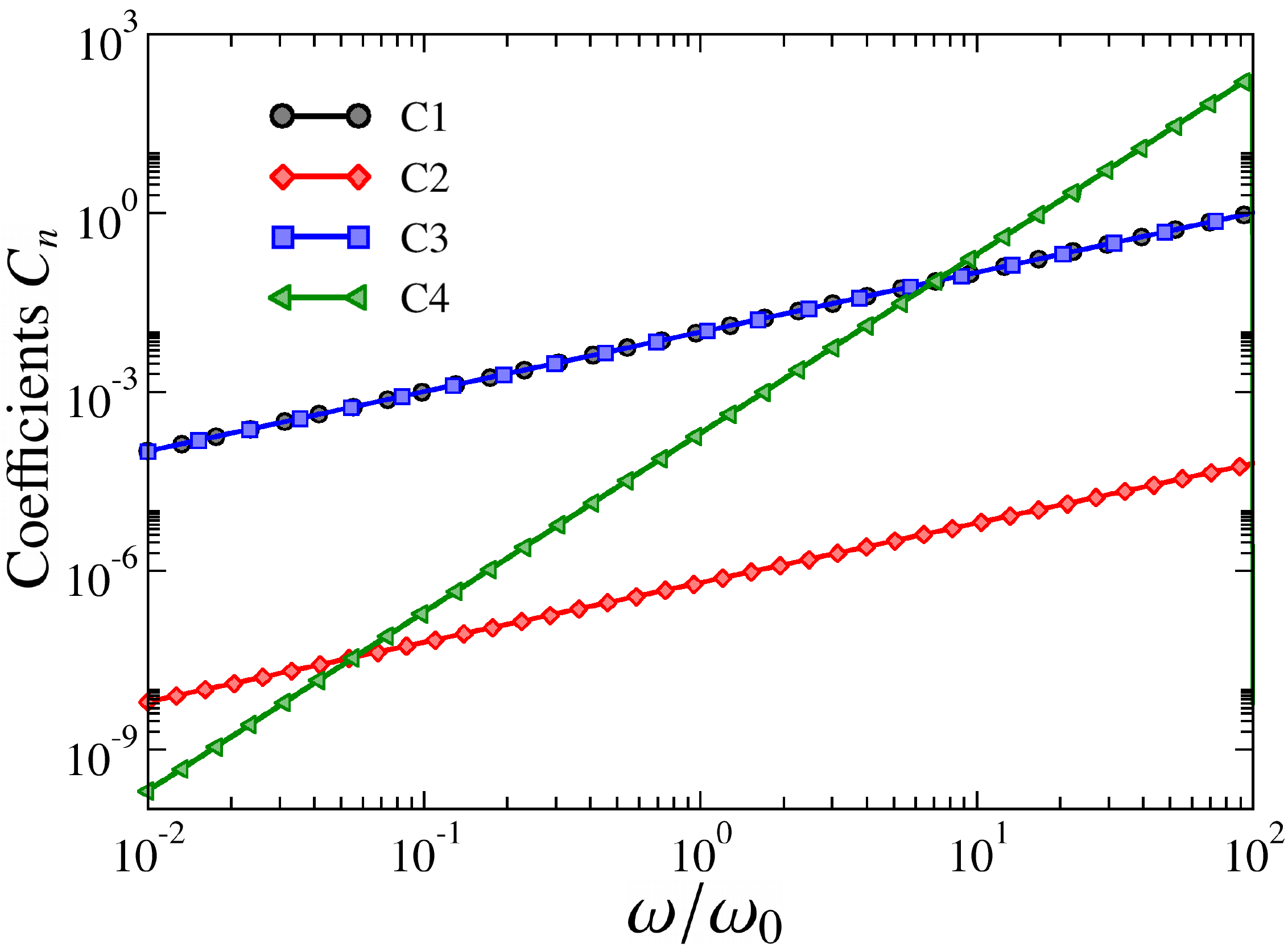}\label{Non-Osci}}\quad
	\subfloat[Non-inertial frame]{\includegraphics[scale=0.25]{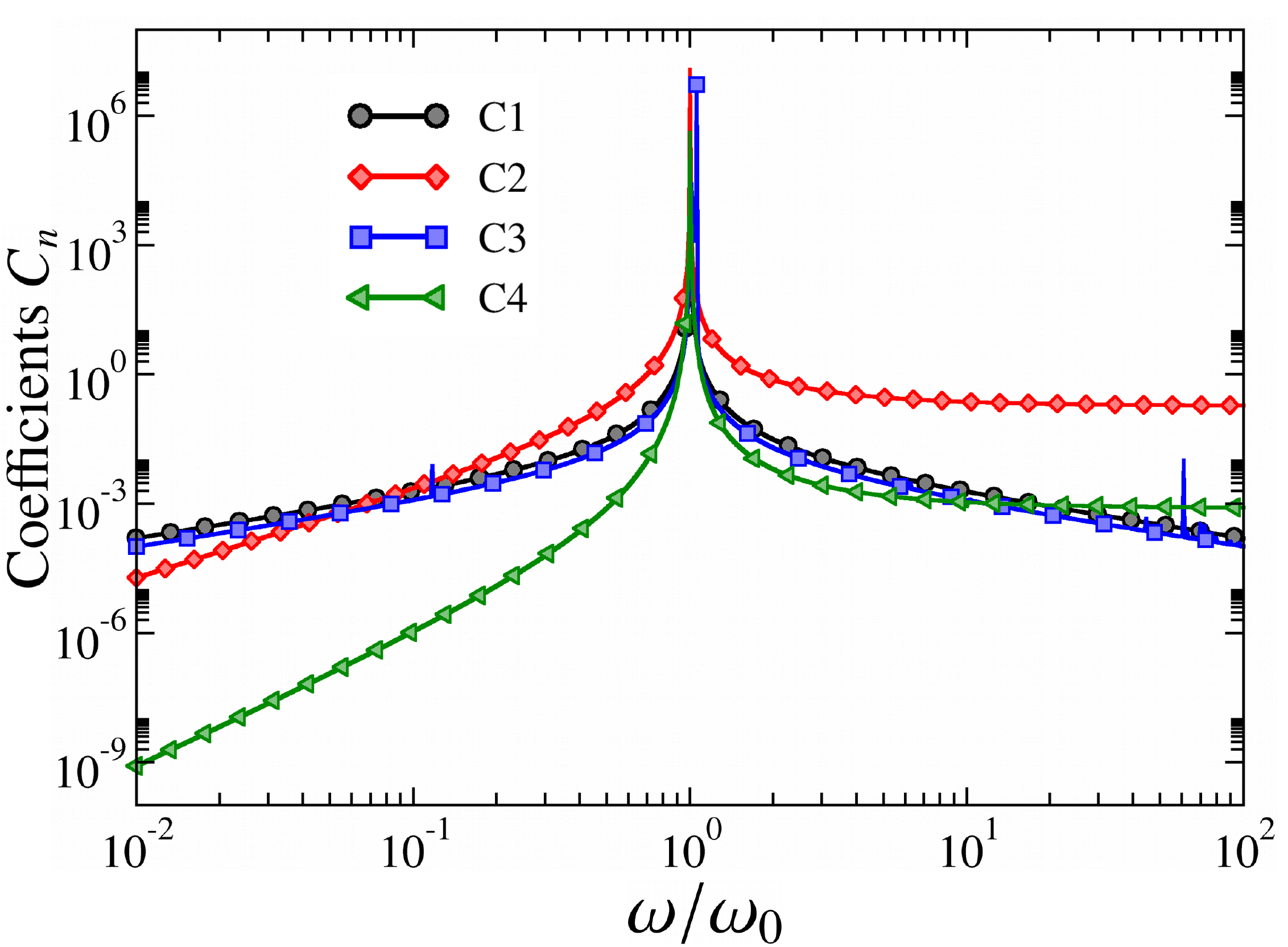}\label{Osci}}
	\caption{\eqref{Fig1} Theoretical and experimental fidelities for the quantum dynamics. The symbols and lines represent experimental data and theoretical results, respectively. The error bars are obtained from 60,000 binary-valued measurements for each data point and are not larger than 1.6\%. \eqref{Non-Osci} We show the coefficients $C_{n}$ as function of $\omega/\omega_{0}$ computed for the inertial frame. \eqref{Osci} We show the coefficients $C_{n}$ as function of $\omega/\omega_{0}$ computed for the non-inertial frame. We set $\omega_{0} = 2\pi \times 1.0$~MHz, $\omega_{\text{T}} = 2\pi \times 20.0$~KHz, and $\tau=100~\mu$s.}
\end{figure*}

In Fig.~\ref{Fig1}, we experimentally compute the fidelity of obtaining the system in the $\ket{\psi_{\text{ad}}(t)}$, where we use the fidelity as $\Fcal(t) = |\text{Tr}[\rho(t)\rho_{\text{ad}}(t)]|$, where $\rho(t)$ is solution of the Eq.~\eqref{VonNeuEq} and $\rho_{\text{ad}}(t) = \ket{E_{1}(t)}\bra{E_{1}(t)}$. We show the experimental results for three different situations, where we have $\omega \gg \omega_{0}$, $\omega \ll \omega_{0}$ and $\omega \approx \omega_{0}$. When we have $|\omega - \omega_{0}| \gg |\omega_{\text{T}}|$, the condition should provide us $C_{n}\ll1$. However, looking at Fig.~\ref{Non-Osci} we can see that such result is not obtained in case $\omega \gg \omega_{0}$. Therefore, all the ACs provided by  Eq.~\eqref{Conditions} are not necessary, once we have adiabaticity even in case where the ACs are not obeyed. On the other hand, in the near-to-resonance situation we have no adiabaticity (once the fidelity is much smaller than $1$), in contrast with Fig.~\ref{Non-Osci}, where we get $C_{n} \lesssim 10^{-2}$. Therefore, the ACs are not sufficient for studying the adiabatic behavior of our system. In conclusion, rather differently from the system considered in Ref. \cite{Suter:08}, all the ACs provided by Eq.~\eqref{Conditions} are not applicable to the dynamics governed by the Hamiltonian in Eq.~\eqref{ResH}. These results imply that a direct
application of the ACs yields neither sufficient nor necessary.

At this point, providing a new condition for adiabaticity could be a natural path to follow. Nevertheless, in order to investigate the applicability of ACs, we will implement, similarly as in classical mechanics, a transformation to a non-inertial frame in Schr\"odinger equation. By considering frame representation in quantum mechanics, Eq.~\eqref{VonNeuEq} can be taken as Schr\"odinger equation in an inertial frame~\cite{Klink:97}. To introduce a non-inertial frame, we can perform a rotation using the unitary time-dependent operator $\Ocal(t) = e^{i \omega t \sigma_{z}}$. In this frame, the dynamics is given by
\begin{equation}
\dot{\rho}_{\Ocal}(t) = (-i/\hbar)[H_{\Ocal}(t) ,\rho_{\Ocal}(t)] \text{ , } \label{RotEq}
\end{equation}
where $H_{\Ocal}(t) = \Ocal(t)H(t)\Ocal^{\dagger}(t)+ i\hbar \dot{\Ocal}(t)\Ocal^{\dagger}(t)$ and $\rho_{\Ocal}(t) = \Ocal(t)\rho(t)\Ocal^{\dagger}(t)$. The
contribution $i\hbar \dot{\Ocal}(t)\Ocal^{\dagger}(t)$ in $H_{\Ocal}(t) $ can be interpreted as a ``fictitious potential"~\cite{Klink:97}. This procedure is a common strategy, e.g., in nuclear magnetic resonance, where we use the non-inertial frame to describe the system dynamics~\cite{Sarthour:Book,Nielsen:Book}. By computing the non-inertial Hamiltonian $H_{\Ocal}(t)$ we find
$H_{\Ocal}(t) = (\omega_{0} - \omega) \sigma_{z}/2 + \sin ( \omega t) \tan \theta \vec{\omega}_{xy}(t)\cdot\vec{\sigma}_{xy}$, 
with $\vec{\omega}_{xy}(t) = \omega_{0}[\cos(\omega t) \hat{x} - \sin(\omega t) \hat{y}]/2$ and $\vec{\sigma}_{xy} = \sigma_{x} \hat{x} + \sigma_{y} \hat{y}$. Now, if we compute the conditions $C_{n}$ considering the set of eigenstate and energies of the new Hamiltonian $H^{\prime}(t)$ we obtain the curves shown in Fig.~\ref{Osci}. Thus, considering the results in Figs.~\ref{Fig1} and~\ref{Osci}, it is possible to conclude that the coefficients $C_{n}$ computed in the non-inertial frame allow us to successfully describe the adiabaticity of the inertial frame. This is in contrast with previous results, which indicated that the ACs may be problematic as we consider oscillating or rotating fields in resonant conditions~\cite{Duki:05,Amin:09}. In particular, notice that even the traditional AC, when analyzed in this non-inertial frame, becomes \textit{sufficient} and \textit{necessary} for the adiabatic behavior of the single-qubit oscillating Hamiltonian in Eq.~(\ref{ResH}).

\emph{Validation mechanism for ACs and frame-dependent adiabaticity --}
We now establish a general validation mechanism for ACs connecting inertial and non-inertial frames, with special focus on cases under resonant conditions.
This approach is applicable beyond the
single-qubit system previously considered, holding for more general multi-particle quantum systems.
Let us consider a Hamiltonian $H(t)$ in an inertial reference frame and its non-inertial counterpart
$H_{\Ocal}(t) $, where the change of reference frame is provided by a generic time-dependent unitary $\Ocal(t)$.
The Hamiltonians $H(t)$ and $H_{\Ocal}(t)$ obey eigenvalue equations given by
$ H(t) |E_n(t) \rangle = E_n(t) |E_n(t) \rangle$ and $ H_{\Ocal}(t) |E^{\Ocal}_n(t) \rangle = E^{\Ocal}_n(t) |E^{\Ocal}_n(t) \rangle$,
with $[H(t),H(t')]\neq0$ and $[H_{\Ocal}(t),H_{\Ocal}(t')]\neq0$, in general.
The adiabatic dynamics in the inertial frame, which is governed by $H(t)$, can be defined through its corresponding evolution operator
$U(t,t_{0})=\sum_{n} e^{i \int _{t_{0}}^{t}\theta_{n}(\xi)d\xi}\ket{E_{n}(t)}\bra{E_{n}(t_{0})}$,
where $\theta_{n}(t) = -E_n(t)/\hbar + i \langle E_n(t) | (d/dt) | E_n(t)\rangle$ is the adiabatic phase, which composed by its dynamic and
geometric contributions, respectively.
Then, we can connect the adiabatic evolution in the inertial and non-inertial frames through the
theorem below.
\begin{theorem}\label{Theorem1}
Consider a Hamiltonian $H(t)$ and its non-inertial counterpart $H_{\Ocal}(t)=\Ocal(t)H(t)\Ocal^{\dagger}(t) + i\hbar \dot{\Ocal}(t)\Ocal^{\dagger}(t)$,
with $\Ocal(t)$ an arbitrary unitary transformation. The eigenstates of $H(t)$ and $H_{\Ocal}(t)$ are denoted by $\ket{E_{k}(t)}$ and $\ket{E^{\Ocal}_{m}(t)}$,
respectively. Then, if a quantum system ${\cal S}$ is prepared at time $t=t_0$ in a particular eigenstate $\ket{E_{k}(t_{0})}$ of $H(t_0)$, then the adiabatic evolution
of ${\cal S}$ in the inertial frame, governed by $H(t)$, is associated with the adiabatic evolution of ${\cal S}$ in the non-inertial frame, governed by $H_{\Ocal}(t)$,
if and only if
\begin{eqnarray}
|\bra{E^{\Ocal}_{m}(t)}\Ocal(t)\ket{E_{k}(t)}| = |\bra{E^{\Ocal}_{m}(t_{0})}\Ocal(t_{0})\ket{E_{k}(t_{0})}|  \,\,\,\,\, \forall t,m  \,\,\, \text{ , }\label{condT1}
\end{eqnarray}
where $t \in [t_{0},\tau]$, with $\tau$ denoting the total time of evolution.
Conversely, if the adiabatic dynamics in the non-inertial frame starts in $\ket{E^{\Ocal}_{m}(t_{0})}$, then the dynamics in the
inertial frame is also adiabatic if and only if Eq.~(\ref{condT1}) is satisfied.
\end{theorem}

The proof is provided in the Supplementary Material. Notice that Theorem~\ref{Theorem1} establishes that,
if Eq.~(\ref{condT1}) is satisfied, then a non-adiabatic behavior in the non-inertial frame ensures a non-adiabatic behavior in the original frame and vice-versa,
provided that the evolution starts in a single eigenstate of the initial Hamiltonian. Then, we can apply this result to general resonant Hamiltonians.
A typical scenario exhibiting resonance phenomena appears when a physical system is coupled to both a static high intensity field $\vec{B}_{0}$
and a time-dependent transverse field $\vec{B}_{\text{T}}(t)$, where $||\vec{B}_{\text{T}}(t)|| \ll ||\vec{B}_{0}||$.
Here, we will consider that the transverse field $\vec{B}_{\text{T}}(t)$ is associated to a single rotating or oscillating field with frequency $\omega$.
In this context, we can write a general multi-qubit Hamiltonian as
\begin{equation}
H(\omega,t) = \hbar \omega_{0} H_{0} + \hbar \omega_{\text{T}} H_{\text{T}}(\omega,t) \text{ , } \label{HOrig}
\end{equation}
where the contributions $\hbar \omega_{0} H_{0}$ and $\hbar \omega_{\text{T}} H_{\text{T}}(\omega,t)$ depend on the fields $\vec{B}_{0}$ and $\vec{B}_{\text{T}}(t)$, respectively. Since $\vec{B}_{0} \perp \vec{B}_{\text{T}}(t)$, we observe that $[H_{\text{T}}(\omega,t),H_{0}]~\neq~0$. In the case $||\vec{B}_{\text{T}}(t)||~\ll~||\vec{B}_{0}||$, the eigenstates $\ket{E_{n}(t)}$ of the Hamiltonian $H(\omega,t)$ can be written as $\ket{E_{n}(t)} \approx \ket{E^{0}_{n}}$, where $\ket{E^{0}_{n}}$ is a stationary eigenstate of the Hamiltonian $\hbar \omega_{0} H_{0}$.

If we have a far-from-resonance situation, we can approximate the dynamics obtained from $H(\omega,t)$ as that one driven by $\hbar \omega_{0} H_{0}$. However, in a near-to-resonance field configuration the most convenient way to study the system dynamics is by adopting a change of reference frame. A general approach to frame change is obtained by the choice $\Ocal(\omega,t) = e^{i \omega H_{0} t }$. Then, from Eq.~\eqref{HOrig}, we can show that
\begin{eqnarray}
H_{\Ocal}(\omega,t) = \hbar \left( \omega_{0} - \omega \right) H_{0} + \hbar \omega_{\text{T}} H_{\Ocal , \text{T}}(\omega,t) \text{ . }
\end{eqnarray}
where $H_{\Ocal,\text{T}}(\omega,t) = \Ocal(t)H_{\text{T}}(t)\Ocal^{\dagger}(t)$. It is worth mentioning that $[H_{\Ocal,\text{T}}(\omega,t),H_{0}]~\neq~0$, once $[H_{\text{T}}(\omega,t),H_{0}]~\neq~0$. In addition, since $H_{\Ocal,\text{T}}(\omega,t)$ is constrained to $H_{\text{T}}(t)$ through a unitary transformation, $||H_{\text{T}}(\omega,t)|| = ||H_{\Ocal,\text{T}}(\omega,t)||$. Therefore, due to the quantity $ \omega_{0} - \omega $ in the first term of $H_{\Ocal}{(\omega,t)}$, the contribution of
$H_{\text{T}}(\omega,t)$ cannot be ignored in this new frame.

As shown in the Supplementary Material, by considering the generic Hamiltonian in Eq.~\eqref{HOrig}, we obtain that Eq.~(\ref{condT1}) in Theorem~\ref{Theorem1} is
automatically satisfied if the quantum system is in a \textit{far-from resonance} configuration $|\omega - \omega_{0}| \gg |\omega_{\text{T}}|$,
so that the adiabatic dynamics in the inertial frame can be always predicted from the adiabaticity analysis in the non-inertial frame.
For this reason, the curves in Fig.~\ref{Osci} can correctly describe the adiabatic behavior exhibited in Fig.~ \ref{Fig1}, yielding $C_{n} \ll 1$ for $|\omega| \gg |\omega_{0}|$
and $|\omega| \ll |\omega_{0}|$. On the other hand, at resonance (or near-to-resonance) configuration $|\omega - \omega_{0}| \ll |\omega_{\text{T}}|$,
Eq.~(\ref{condT1}) in Theorem 1 reduces to the rather simple condition $|\interpro{E^{\Ocal}_{m}(t)}{E^{0}_{k}}| = |\interpro{E^{\Ocal}_{m}(t_{0})}{E^{0}_{k}}|$.
Hence, provided a generic Hamiltonian given by Eq.~(\ref{HOrig}) at resonance (or near-to-resonance) situation,
if the corresponding Hamiltonian in the non-inertial frame has \textit{time-dependent} eigenstates obeying
$|\interpro{E^{\Ocal}_{m}(t)}{E^{0}_{k}}| = \text{constant}$, $\forall t,m$, for a particular initial state $|{E^{0}_{k}}\rangle$, then a
non-adiabatic evolution in the non-inertial frame implies in non-adiabatic evolution in the inertial frame. This is exactly
the case for the Hamiltonian in Eq. \eqref{ResH}, with the violation of adiabaticity at resonance illustrated in Fig.~\ref{Osci}
for all the ACs considered.

\emph{Revisiting the problem of the spin-1/2 particle in a rotating magnetic field --} We now apply our general treatment to the
NMR Hamiltonian discussed by Du \textit{et al.}~\cite{Suter:08}. The dynamics describes a single spin-1/2 particle coupled to a static field
$\vec{B}_{\text{0}} = B_{\text{0}}\hat{z}$ and a transverse radio-frequency field
$\vec{B}_{\text{rf}}(t) = B_{\text{rf}}[\cos (\omega) \hat{x} + \sin (\omega) \hat{y}]$, with Hamiltonian given by
\begin{equation}
H_{\text{nmr}} ( t ) = (\omega _{0}/2)\sigma _{z}+ (\omega _{\text{rf}}/2)\left[ \cos ( \omega t ) \sigma
_{x}+\sin ( \omega t ) \sigma _{y}\right] \text{ ,}
\label{SuterHamiltonian}
\end{equation}
where $|\omega _{0}|\gg|\omega _{\text{rf}}|$. The system is prepared in an eigenstate of $\sigma_z$ and the frequencies are chosen
such that the standard AC is satisfied~\cite{Suter:08}. In this scenario, the violations and agreements about ACs for this system have widely
been discussed in literature~\cite{Tong:10,Comments-Tong-Comparat:11,Comments-Tong-Zhao:11,Reply-Tong:11}. Here we analyze this
Hamiltonian from a different point of view. By writing the system dynamics in the non-inertial frame through
$\Ocal(t) = e^{\frac{i}{\hbar }\frac{\omega }{2}t\sigma _{z}}$, we obtain
$H_{\Ocal}^{\text{nmr}} = (\omega _{0}-\omega)\sigma
_{z}/2+(\omega _{\text{rf}}/2)\sigma _{x}$.
Since this Hamiltonian is time-independent, the dynamics under $H_{\Ocal}^{\text{nmr}}$ is trivially adiabatic, with all ACs in Eq.~\eqref{Conditions} satisfied.
Therefore, the is no direct visualization of the resonant point. However, Theorem 1 cannot be directly applied here near to resonance because the initial
state in this case is not an individual eigenstate of $H_{\Ocal}^{\text{nmr}}$,
since $H_{\Ocal}^{\text{nmr}}$ is approximately proportional to $\sigma_x$. 
We can circumvent this problem by taking advantage of the time-independence
of $H_{\Ocal}^{\text{nmr}}$. More specifically, we start from the evolution operator $U_{\Ocal}(t,t_{0}) = e^{-\frac{i}{\hbar} H_{\Ocal}(t-t_{0})}$ in the non-inertial frame
and investigate under which conditions we may obtain an adiabatic dynamics in the inertial frame. This can be suitably addressed by Theorem 2 below.
\begin{theorem} \label{Theorem2}
Consider a Hamiltonian $H(t)$ and its non-inertial counterpart $H_{\Ocal}=\Ocal(t)H(t)\Ocal^{\dagger}(t) + i\hbar \dot{\Ocal}(t)\Ocal^{\dagger}(t)$,
with $\Ocal(t)$ an arbitrary unitary transformation and $H_{\Ocal}$ a constant Hamiltonian.
The eigenstates of $H(t)$ and $H_{\Ocal}$ are denoted by $\ket{E_{k}(t)}$ and $\ket{E^{\Ocal}_{m}}$,
respectively. Then, if a quantum system ${\cal S}$ is prepared at time $t=t_0$ in a particular eigenstate $\ket{E_{n}(t_{0})}$ of $H(t_0)$,
then the adiabatic evolution of ${\cal S}$ in the inertial frame, governed by $H(t)$, occurs if and only if
\begin{equation}
| \bra{E_{k}(t)}U_{\Ocal}(t,t_{0})\ket{E_{n}(t_{0})} | = |\interpro{E_{k}(t_{0})}{E_{n}(t_{0})}|  \,\,\,\,\, \forall t,k  \,\,\, \text{ , } \label{ApU2}
\end{equation}
where $t \in [t_{0},\tau]$, with $\tau$ denoting the total time of evolution, and
$U_{\Ocal}(t,t_{0}) = \Ocal^{\dagger}(t)e^{-\frac{i}{\hbar} H_{\Ocal}(t-t_{0})}\Ocal(t_{0})$.
\end{theorem}
The proof is provided in the Supplementary Material. The experimental results in Ref.~\cite{Suter:08} can be validated by Theorem 2, since
the Hamiltonian in Eq.~\eqref{SuterHamiltonian} satisfies Eq.~(\ref{ApU2}) in a far-from resonance situation and
violates it at resonance. In fact, the initial state $|\psi(0)\rangle$ can be approximately written as $|\psi(0)\rangle = \ket{E_{n}(0)} \approx |n\rangle$ [with $\sigma_z |n\rangle =(-1)^{(n+1)} |n\rangle$]. Thus, Eq. \eqref{ApU2} provides the condition $| \bra{k}e^{-\frac{i}{\hbar} H_{\Ocal}t}\ket{n} | = |\interpro{k}{n}| = \delta_{kn}$, $\forall k$ and $\forall t\in [0,\tau]$.
In a far-from-resonance situation, we have
$H_{\Ocal}^{\text{nmr}} \approx \frac{\omega _{0}-\omega }{2}\sigma_{z}$, and we conclude that $| \bra{k}e^{-\frac{i}{\hbar} H_{\Ocal}t}\ket{n} | \approx \delta_{kn}$. This shows that the dynamics in the inertial frame is (approximately) adiabatic far from resonance. On the other hand, near to resonance, we get $H_{\Ocal}^{\text{nmr}} \approx \frac{\omega _{\text{rf}}}{2}\sigma_{x}$, where we can immediately conclude that
$| \bra{k}e^{-\frac{i}{\hbar} H_{\Ocal}t}\ket{n} | \napprox \delta_{kn}$ for any $t\in [0,\tau]$.

\emph{Conclusions --} We have introduced a framework to validate ACs in generic discrete multi-particle Hamitonians, which is rather convenient to
analyze quantum systems at resonance. This is based on the analysis of ACs in a suitably designed non-inertial reference frame. In particular,
we have both theoretically and experimentally shown that several relevant ACs [provided by Eq.~\eqref{Conditions}], which include the traditional AC, are sufficient
and necessary to describe the adiabatic behavior of a qubit in an oscillating field given by Eq. (\ref{ResH}). In this case, sufficiency and necessity are fundamentally
obtained through the non-inertial frame map, with all the conditions failing to point out the adiabatic behavior in the original reference frame.
The experimental realization has been
performed through a single trapped Ytterbium ion, with excellent agreement with the theoretical results.
More generally, the validation of ACs has been expanded to arbitrary Hamiltonians through
Theorems 1 and 2, with detailed conditions provided for a large class of Hamiltonians in the form of Eq.~\eqref{HOrig}.
Therefore, instead of looking for new approaches for defining ACs, we have introduced a mechanism based on ``fictitious potentials" (associated with non-inertial frames)
to reveal a correct indication of ACs, both at resonance and off-resonant situations. In addition, as a further example, we discuss how the
validation mechanism through non-inertial frames can be useful to describe the results presented in Ref. \cite{Suter:08}, where the adiabatic dynamics
of a single spin-$1/2$ in NMR had been previously investigated. More general settings, such as decoherence effects,
are left for future research.

\emph{Acknowledgments --} We thank Yuan-Yuan Zhao for valuable discussion. This work was supported by the National Key Research and Development Program of China (No. 2017YFA0304100), National Natural Science Foundation of China (Nos. 61327901, 61490711, 11774335, 11734015, 11474268, 11374288, 11304305), Anhui Initiative in Quantum Information Technologies (AHY070000, AHY020100),  Anhui Provincial Natural Science Foundation (No. 1608085QA22), Key Research Program of Frontier Sciences, CAS (No. QYZDY-SSWSLH003), the National Program for Support of Top-notch Young Professionals (Grant No. BB2470000005), the Fundamental Research Funds for the Central Universities (WK2470000026). A.C.S. is supported by Conselho Nacional de Desenvolvimento Cient\'{\i}fico e Tecnol\'ogico (CNPq-Brazil). M.S.S. is supported by CNPq-Brazil (No. 303070/2016-1) and Funda\c{c}\~ao Carlos Chagas Filho de Amparo \`a Pesquisa do Estado do Rio de Janeiro (FAPERJ) (No. 203036/2016). A.C.S., F.B., and M.S.S. also acknowledge
financial support in part by the Coordena\c{c}\~ao de Aperfei\c{c}oamento de Pessoal de N\'{\i}vel Superior - Brasil (CAPES) (Finance Code 001) and by the Brazilian
National Institute for Science and Technology of Quantum Information (INCT-IQ).


\newpage

\hspace{10cm}

\newpage

\appendix

\renewcommand{\thesection}{S-\arabic{section}}
\renewcommand{\theequation}{S\arabic{equation}}
\setcounter{equation}{0}  
\renewcommand{\thefigure}{S\arabic{figure}}

\onecolumngrid

\vspace*{0.4cm}
\begin{center}
	{\large \bf Supplementary Material for:}
	\vskip 0.2 cm
	{\large \bf Validation of Quantum Adiabaticity through Non-Inertial Frames and Its Trapped-Ion Realization} \\
	\vspace{0.6cm}
	Chang-Kang Hu,$^{1,2}$ Jin-Ming Cui,$^{1,2,{\color{blue}\ast}}$ Alan C. Santos,$^{3,{\color{blue}\dagger}}$ \\
	Yun-Feng Huang,$^{1,2,{\color{blue}\ddagger}}$ Chuan-Feng Li,$^{1,2{\color{blue}\ast\ast}}$ Guang-Can Guo,$^{1,2}$ Frederico Brito,$^{4,{\color{blue}\S}}$ Marcelo S. Sarandy,$^{3,{\color{blue}\P}}$  \\
	$^1${\small \it CAS Key Laboratory of Quantum Information, University of Science and Technology of China, Hefei, 230026, People’s Republic of China} \\
	$^2${\small \it Synergetic Innovation Center of Quantum Information and Quantum Physics,\\University of Science and Technology of China, Hefei, 230026, People’s Republic of China} \\
	$^3${\small \it Instituto de F\'{i}sica, Universidade Federal Fluminense, \\Av. Gal. Milton Tavares de Souza s/n, Gragoat\'{a}, 24210-346 Niter\'{o}i, Rio de Janeiro, Brazil}\\
	$^4${\small \it Instituto de Física de São Carlos, Universidade de São Paulo, C.P. 369, 13560-970 São Carlos, SP, Brazil}\\
	\vspace{0.1cm}
	$^{\color{blue}\ast}${\small jmcui@ustc.edu.cn} \quad $^{\color{blue}\dagger}${\small ac\_santos@id.uff.br} \quad $^{\color{blue}\ddagger}${\small hyf@ustc.edu.cn} \quad $^{\color{blue}{\ast\ast}}${\small cfli@ustc.edu.cn} \quad $^{\color{blue}\S}${\small fbb@ifsc.usp.br} \quad $^{\color{blue}\P}${\small msarandy@id.uff.br}

\end{center}
\vspace{0.6cm}

\twocolumngrid

\section*{Proof of Theorem 1}

Let us consider two Hamiltonians, an inertial frame Hamiltonian $H(t)$ and its non-inertial counterpart $H_{\Ocal}(t)$, which are related by a time-dependent unitary $\Ocal(t)$. The dynamics associated with Hamiltonians $H(t)$ and $H_{\Ocal}(t)$ are given by
\begin{eqnarray}
\dot{\rho}(t) &=& \frac{1}{i\hbar}[H(t),\rho(t)] \text{ , } \label{ApSchodingerEq} \\
\dot{\rho}_{\Ocal}(t) &=& \frac{1}{i\hbar}[H_{\Ocal}(t),\rho_{\Ocal}(t)] \text{ , } \label{ApSchodingerEqRot}
\end{eqnarray}
where $H_{\Ocal}(t) = \Ocal(t)H(t)\Ocal^{\dagger}(t) + i\hbar \dot{\Ocal}(t)\Ocal^{\dagger}(t) $ and $\rho_{\Ocal}(t) = \Ocal(t)\rho(t)\Ocal^{\dagger}(t)$. Then, the connection between the evolved states $\ket{\psi(t)}$ and $\ket{\psi_{\Ocal}(t)}$ in inertial and non-inertial frames, respectively, is given by $\ket{\psi_{\Ocal}(t)} = \Ocal(t)\ket{\psi(t)}$, $\forall t \in [t_0,\tau]$. By considering the initial state in inertial frame given by a single eigenstate of $H(t)$, namely \st{$c$} $\ket{\psi(t_0)}=\ket{E_{k}(t_0)}$, the adiabatic dynamics in this frame is written as
\begin{eqnarray}
\ket{\psi(t)} = e^{i\int_{t_{0}}^{t} \theta_{k}(\xi)d\xi}\ket{E_{k}(t)} \text{ , }
\end{eqnarray}
where $\theta_{k}(t) = -E_k(t)/\hbar + i \langle E_k(t) | (d/dt) | E_k(t)\rangle$ is the adiabatic phase composed by the dynamical and geometrical phase, respectively. On the other hand, an adiabatic behavior is obtained in non-inertial frame if and only if
\begin{eqnarray}
|\interpro{E^{\Ocal}_{m}(t)}{\psi_{\Ocal}(t)}| = |\interpro{E^{\Ocal}_{m}(t_{0})}{\psi_{\Ocal}(t_{0})}| \text{ , } \forall m \, , \,  \forall t \in [t_0,\tau] \text{ . }
\end{eqnarray}
Therefore, we can write
\begin{eqnarray}
|\interpro{E^{\Ocal}_{m}(t)}{\psi_{\Ocal}(t)}| &=& |\interpro{E^{\Ocal}_{m}(t_{0})}{\psi_{\Ocal}(t_{0})}| ,  \nonumber \\
|\bra{E^{\Ocal}_{m}(t)}\Ocal(t)\ket{\psi(t)}| &=& |\bra{E^{\Ocal}_{m}(t_{0})}\Ocal(t_{0})\ket{\psi(t_{0})}| , \nonumber \\
|\bra{E^{\Ocal}_{m}(t)}\Ocal(t)\ket{E_{k}(t)}| &=& |\bra{E^{\Ocal}_{m}(t_{0})}\Ocal(t_{0})\ket{E_{k}(t_{0})}| \text{ . } \label{ApCond}
\end{eqnarray}
Thus, Eq.~(\ref{ApCond})  establishes a necessary and sufficient condition to obtain an adiabatic evolution in the non-inertial frame,
assuming an adiabatic evolution in the original frame.
To conclude our proof, let us consider the converse case, where the system starts in a eigenstate of $\ket{E^{\Ocal}_{m}(t_{0})}$ in non-inertial frame. If the dynamics is adiabatic we write
\begin{eqnarray}
\ket{\psi_{\Ocal}(t)} = e^{i\int_{t_{0}}^{t} \theta^{\Ocal}_{m}(\xi)d\xi}\ket{E^{\Ocal}_{m}(t)} \text{ , }
\end{eqnarray}
where $\theta^{\Ocal}_{m}(t)$ is the adiabatic phase collected in this frame. The dynamics will be adiabatic in the inertial frame if and only if
\begin{eqnarray}
|\interpro{E_{m}(t)}{\psi(t)}| = |\interpro{E_{m}(t_{0})}{\psi(t_{0})}| \text{ , } \forall m \, , \,  \forall t \in [t_0,\tau] \text{ . }
\end{eqnarray}

Therefore, by using the same procedure as before, we get the condition
\begin{eqnarray}
|\bra{E_{k}(t)}\Ocal^{\dagger}(t)\ket{E^{\Ocal}_{m}(t)}| &=& |\bra{E_{k}(t_{0})}\Ocal^{\dagger}(t_{0})\ket{E^{\Ocal}_{m}(t_{0})}| \text{ , }
\end{eqnarray}
which is equivalent to Eq.~\eqref{ApCond}. This ends the proof of Theorem 1.

\section*{Application of Theorem 1 to the Hamiltonian in Eq. \eqref{HOrig}}
Let us consider a generic system under action of a single time-dependent oscillating/rotating field with characteristic
frequency $\omega$, whose Hamiltonian reads
\begin{eqnarray}
H(\omega,t) = \hbar \omega_{0} H_{0} + \hbar \omega_{\text{T}} H_{\text{T}}(\omega,t) \text{ , } \label{ApHOrig}
\end{eqnarray}
where we consider the transverse term $\hbar \omega_{\text{T}} H_{\text{T}}(\omega,t)$ as a perturbation, so that $||\omega_{0} H_{0}|| \gg ||\omega_{\text{T}} H_{\text{T}}(\omega,t)||$, $\forall t\in [0,\tau]$. In this case, the eigenstates $\ket{E_{n}(t)}$ and energies $E_{n}(t)$ of $H(\omega,t)$ can be obtained as perturbation of eigenstates $\ket{E_{n}^{0}}$ and energies $E_{n}^{0}$ of $\hbar \omega_{0} H_{0}$ as (up to a normalization coefficient)
\begin{eqnarray}
\ket{E_{n}(t)} &=& \ket{E_{n}^{0}} + \Ocalb \left(||\hbar \omega_{\text{T}} H_{\text{T}}(\omega,t)|| \right) \text{ , } \label{ApvecnonRot}\\
E_{n}(t) &=& E_{n}^{0} + \Ocalb \left(||\hbar \omega_{\text{T}} H_{\text{T}}(\omega,t)|| \right) \text{ . } \label{ApEnenonRot}
\end{eqnarray}
On the other hand, in the non-inertial frame, we have $H_{\Ocal}(t) = \Ocal(t)H(t)\Ocal^{\dagger}(t) + i\hbar \dot{\Ocal}(t)\Ocal^{\dagger}(t) $, which yields
\begin{eqnarray}
H_{\Ocal}(\omega,t) = \hbar \left( \omega_{0} - \omega \right) H_{0} + \hbar \omega_{\text{T}} H_{\Ocal , \text{T}}(\omega,t) \text{ . } \label{ApHrot}
\end{eqnarray}
where $H_{\Ocal , \text{T}}(\omega,t) = \Ocal(t)H_{\text{T}}(\omega,t)\Ocal^{\dagger}(t)$. Now, we separately consider two specific cases:

$\bullet$ \emph{Far-from resonance situation $|\omega_{0} - \omega|\gg |\omega_{\text{T}}|$}: In this case, the term $\hbar \omega_{\text{T}} H_{\Ocal , \text{T}}(\omega,t)$ in Eq. \eqref{ApHrot} works as a perturbation. Therefore the set of eigenvectors of $H_{\Ocal}(\omega,t)$ reads
\begin{eqnarray}
\ket{E^{\Ocal}_{n}(t)} &=& \ket{E_{n}^{0}} + \Ocalb \left(||\hbar \omega_{\text{T}} H_{\text{T}}(\omega,t)|| \right) \text{ , } \label{ApvecRot}
\end{eqnarray}
where we have used that the energy gaps $\tilde{E}_{n}^{0} - \tilde{E}_{k}^{0}$ of the Hamiltonian $\hbar \left( \omega_{0} - \omega \right) H_{0}$ are identical to energy gaps $E_{n}^{0} - E_{k}^{0}$ of $\hbar \omega_{0} H_{0}$ and $||\hbar \omega_{\text{T}} H_{\Ocal , \text{T}}(\omega,t)|| = ||\hbar \omega_{\text{T}} H_{\text{T}}(\omega,t)||$. Thus,
from Eqs. \eqref{ApvecnonRot} and \eqref{ApvecRot} we conclude, for any eigenstate  $\ket{E_{k}{(t)}}$, that
\begin{eqnarray}
\bra{E^{\Ocal}_{m}(t)}\Ocal(t)\ket{E_{k}(t)} \approx e^{i\frac{\omega}{\omega_{0}} \frac{E_{k}^{0}}{\hbar} t}\delta_{mk} \text{ , }
\end{eqnarray}
so that we get $|\bra{E^{\Ocal}_{m}(t)}\Ocal(t)\ket{E_{k}(t)}| = \text{constant}$, $\forall m$, $\forall t\in [t_{0},\tau]$.

$\bullet$ \emph{Resonance situation $|\omega_{0} - \omega| \ll |\omega_{\text{T}}|$}: Now, we have a more subtle situation.
Firstly, we can use Eqs. \eqref{ApvecnonRot} and \eqref{ApEnenonRot} to write
\begin{eqnarray}
\Ocal (t)\ket{E_{n}(t)} & = & e^{i\frac{\omega}{\omega_{0}} \frac{E_{n}^{0}}{\hbar}t } \ket{E_{n}^{0}} + \Ocalb \left(||\hbar \omega_{\text{T}} H_{\text{T}}(\omega,t)|| \right) \text{ , } \\
\int_{t_{0}}^{t}\theta_{n}(\xi)d\xi & = & -\frac{E_{n}^{0}}{\hbar}(t-t_{0}) + \Ocalb \left(||\hbar \omega_{\text{T}} H_{\text{T}}(\omega,t)|| \right)  \text{ , }
\end{eqnarray}
so that
\begin{eqnarray}
\bra{E^{\Ocal}_{m}(t)}\Ocal(t)\ket{E_{k}(t)} \approx e^{i\frac{\omega}{\omega_{0}} \frac{E_{k}^{0}}{\hbar} t} \interpro{E^{\Ocal}_{m}(t)}{E^{0}_{k}} \text{ . }
\end{eqnarray}
Now, it is possible to see that if $|\interpro{E^{\Ocal}_{m}(t)}{E^{0}_{k}}| = |\interpro{E^{\Ocal}_{m}(t_{0})}{E^{0}_{k}}|$,  $\forall t\in [t_{0},\tau]$, then we obtain $\left \vert \bra{E^{\Ocal}_{m}(t)}\Ocal(t)\ket{E_{k}(t)} \right \vert = \text{constant}$.

\section*{Proof of Theorem 2}

Let us consider a time-independent Hamiltonian $H_{\Ocal}$ in the non-inertial frame, so that its evolution operator can be written as $U_{\Ocal}(t,t_{0}) = e^{-\frac{i}{\hbar} H_{\Ocal}(t-t_{0})}$. Thus, we can write the dynamics in non-inertial frame as
\begin{eqnarray}
\ket{\psi_{\Ocal}(t)} = e^{-\frac{i}{\hbar} H_{\Ocal}(t-t_{0})}\ket{\psi_{\Ocal}(t_{0})} \text{ . } \label{ApDyn}
\end{eqnarray}
Moreover, assuming adiabatic dynamics in the inertial frame, we get
\begin{eqnarray}
| \interpro{E_{k}(t)}{\psi(t)} | = |\interpro{E_{k}(t_{0})}{\psi(t_{0})}| \text{ . }
\end{eqnarray}
By using the relationship between inertial and non-inertial frames as $\ket{\psi(t)} = \Ocal^{\dagger}(t)\ket{\psi_{\Ocal}(t)}$, we can write
\begin{eqnarray}
| \bra{E_{k}(t)}\Ocal^{\dagger}(t)e^{-\frac{i}{\hbar} H_{\Ocal}(t-t_{0})}\ket{\psi_{\Ocal}(t_{0})} | = |\interpro{E_{k}(t_{0})}{\psi(t_{0})}| \text{ , }
\end{eqnarray}
where we have used the Eq. \eqref{ApDyn}. Now, by taking $\ket{\psi_{\Ocal}(t_{0})} = \Ocal(t_0)\ket{\psi(t_{0})}$, we obtain
\begin{eqnarray}
| \bra{E_{k}(t)}\Ocal^{\dagger}(t)e^{-\frac{i}{\hbar} H_{\Ocal}(t-t_{0})}\Ocal(t_{0})\ket{\psi(t_{0})} | = |\interpro{E_{k}(t_{0})}{\psi(t_{0})}| \label{t2af} \text{ . }
\end{eqnarray}
Thus, by inserting the initial state $\ket{\psi(t_{0})} = \ket{E_{n}(t_{0})}$ in Eq.~(\ref{t2af}), we get
\begin{equation}
| \bra{E_{k}(t)}\Ocal^{\dagger}(t)e^{-\frac{i}{\hbar} H_{\Ocal}(t-t_{0})}\Ocal(t_{0})\ket{E_{n}(t_{0})} | = |\interpro{E_{k}(t_{0})}{E_{n}(t_{0})}| \text{ . }
\end{equation}
This concludes the proof of Theorem 2.

\newpage
\end{document}